\documentclass[amsmath,amssymb]{revtex4}
\usepackage[parfill]{parskip}    
\usepackage{graphicx}
\usepackage{amssymb}
\usepackage{epstopdf}
\usepackage{color}
\usepackage{graphicx}
\usepackage{pdfpages}

\begin{document}

\title{Extended gravitational decoupling in $2+1$ dimensional space--time}
\author{Ernesto Contreras {\footnote{On leave 
from Universidad Central de Venezuela}
\footnote{econtreras@yachaytech.edu.ec}} }
\address{School of Physical Sciences \& Nanotechnology, Yachay Tech University, 100119 Urcuqu\'i, Ecuador\\}
\author{Pedro Bargue\~no}
\address{Departamento de F\'{\i}sica,\\ Universidad de Los Andes, Cra. 1 E No 18 A-10, Bogot\'a, Colombia}
\begin{abstract}
In this work we extend the so--called Minimal Geometric Deformation method in $2+1$ dimensional
space--times with cosmological constant in order to deal with the gravitational decoupling of two circularly symmetric 
sources. We find that, even though the system here studied is lower dimensional and it includes the cosmological constant, the conditions for gravitational decoupling of two circularly symmetric sources coincides with those found in the $3+1$ 
dimensional case.  We obtain that, under certain circumstances, the extended gravitational decoupling leads to the 
decoupling of the sources involved in the sense that both the isotropic and the anisotropic sector  satisfy Einstein's field equations and the final solution corresponds to a non-linear superposition of two metric components.
As particular examples, we implement the method to generate an exterior charged BTZ solution starting from the
BTZ vacuum as the isotropic sector and new $2+1$ black hole solutions imposing a barotropic equation of state
for the anisotropic sector. We also show that the imposition of a polytropic equation of state  of the decoupler matter allows to construct a regular black hole solution in three--dimensional gravity.
\end{abstract}
\maketitle

\section{Introduction}\label{intro}
Needless to say, three dimensional gravity still represents an important source for theoretical 
developments: quantum models \cite{carlip} which
have allowed to get some insight about the nature of quantum gravity in (3+1) dimensions, and  identification of cosmic 
strings solutions with  topological defects in two dimensional condensed matter systems as graphene layers (see, for example, \cite{vozmediano2007a,vozmediano2007b,vozmediano2008}), are some of the examples we can list. 
These are some of the reasons why it is important
to find new solutions, whose interpretation could be influenced by 
the topological nature of 2+1 dimensional gravity. 

Recently, the so called 
Minimal Geometric Deformation (MGD) method, whose original version \cite{ovalle2008}
was developed in the context of the
Randall-Sundrum braneworld \cite{rs1999}, has taught us that the decoupling of sources in General Relativity is a highly non--trivial
problem whose implementation allows to find solutions of Einstein's equations in a broad range of cases
(for more details see, for example, Refs.
\cite{ovalle2009,ovalle2010,casadio2012,ovalle2013,ovalle2013a,
	casadio2014,casadio2015,ovalle2015,casadio2015b,
	ovalle2016, cavalcanti2016,casadio2016a,ovalle2017,
	rocha2017a,rocha2017b,casadio2017a,ovalle2018,ovalle2018bis,
	estrada2018,ovalle2018a,lasheras2018,gabbanelli2018,sharif2018,sharif2018a,sharif2108b,
	fernandez2018,fernandez2018b,
	contreras2018,estrada,contreras2018a,morales,tello18,
	rincon2018,ovalleplb,contreras2018c,contreras2019,contreras2019a,tello2019,lh2019}), including the obtention
of new solutions in $2+1$ dimensions with and without cosmological terms 
\cite{contreras2018,contreras2019,contreras2019a}. 
The method can be summarized as follows. Suppose that we want to solve the Einstein equations for certain 
gravitational source, $T^{(tot)}_{\mu\nu}$. As it is well known, given the nature of the field equations, the problem is, in general, 
highly non--trivial, even in spherical (in $3+1$ dimensions) or circularly (in $2+1$ dimensions) symmetric cases. 
Now let us assume that the source 
$T^{(tot)}_{\mu\nu}$ can be written as a linear combination of two other sources, namely $T_{\mu\nu}^{(m)}$ 
and $\theta_{\mu\nu}$ as
\begin{eqnarray}\label{total}
T^{(tot)}_{\mu\nu}=T_{\mu\nu}^{(m)}+\theta_{\mu\nu}.
\end{eqnarray}
Of course, the general belief is that such a simple splitting of the source does not lead to the separation of the Einstein 
equations for each one of them. However, contrary to what is usually expected, it has been demonstrated that
the MGD method allows to solve Einstein's field equations for each source separately and then, by a suitable superposition, 
to obtain the complete solution of the original source. The remarkable fact here is that the method allows to separate  
Einstein's equations in two sets of simpler differential equations that can be easier to deal with than if we try to solve the 
original system directly. \\
To illustrate how the methods works, in what follows we will describe the MGD decoupling in $2+1$ dimensional space-times .
Let us consider the Einstein field equations \footnote{$\kappa^{2}=8\pi$ in explicit computations.}
\begin{eqnarray}\label{einsorig}
R_{\mu\nu}-\frac{1}{2}R g_{\mu\nu}+\Lambda g_{\mu\nu}=\kappa^{2}T_{\mu\nu}^{tot},
\end{eqnarray}
and assume that the total energy-momentum tensor can be decomposed as Eq. (\ref{total})
with $T^{\mu(m)}_{\nu}=diag(-\rho,p,p)$ and $\theta^{\mu}_{\nu}=diag(-\rho^{\theta},p_{r}^{\theta},p_{\perp}^{\theta})$
representing two non-generic gravitational sources; one of them is a perfect fluid and the second one can be considered as an anisotropic source. In what follows, we shall work with circularly symmetric space--times with a line element parametrized as
\begin{eqnarray}\label{le}
ds^{2}=-e^{\nu}dt^{2}+e^{\lambda}dr^{2}+r^{2}d\phi^{2},
\end{eqnarray}
where $\nu$ and $\lambda$ are functions of the radial coordinate, $r$, only. 
In the context of the MGD approach, a decoupling in the geometric sector can be successfully implemented in terms of a 
decomposition in the radial component of the metric,
namely, 
\begin{eqnarray}\label{decompositon}
e^{-\lambda}=e^{-\mu}+\alpha f,
\end{eqnarray}
where $\mu$, in combination with the metric function $\nu$ of the solution regarding source one, 
is again considered as a solution corresponding to the same source of the 
Einstein field equations and $g$ and $f$ are the geometric deformations undergone by
$\xi$ and $\mu$, ``controlled'' by the free parameter $\alpha$. To be more precise, substituting 
Eq. (\ref{decompositon}) in (\ref{einsorig}) we obtain two sets of differential equations: i) a set of Einstein 
field equations for $\{\nu,\mu\}$ with matter sector given by $T^{\mu(m)}_{\nu}=diag(-\rho,p,p)$, namely

\begin{eqnarray}
\kappa ^2\rho &=& \frac{-2 \Lambda  r-\mu '}{2 r}\label{Oiso1}\\
\kappa ^2 p&=& \frac{2 \Lambda  r+\mu \nu '}{2  r}\label{Oiso2}\\
\kappa ^2 p&=&\frac{4 \Lambda +\mu ' \nu '+\mu \left(2 \nu ''+\nu '^2\right)}{4 }
,\label{Oiso3}
\end{eqnarray}

which is the isotropic source and ii) a set of Einstein field equations 
in sourced by $\theta^{\mu}_{\nu}$ given by

\begin{eqnarray}
\kappa ^2\rho^{\theta}&=&-\frac{\alpha f'}{2 r}\label{Oaniso1}\\
\kappa^{2} p_{r}^{\theta}&=&\frac{\alpha f \nu '}{2 r}\label{Oaniso2}\\
\kappa^{2} p_{\perp}^{\theta}&=&\frac{\alpha f' \nu '+\alpha f \left(2 \nu ''+\nu '^2\right)}{4}.\label{Oaniso3}
\end{eqnarray}

In this sense, given a solution of the 
system $\{\nu,\mu\}$, another solution can be found solving for the second set of equations
involving
the unknowns $\{f,\rho^{\theta},p^{\theta}_{r},p^{\theta}_{\perp}\}$. In spherical ($(3+1)$ dimensional space--times) or 
circularly symmetric space--times ($2+1$ space--time dimensions),
the method leads to three equations for four unknowns. In order to solve the equations, only one
aditional condition has to be implemented. Some of the cases are listed below:
\begin{itemize}
	\item Interior solutions. In this case, an interior solution matched with the
	exterior Schwarzschild vacuum (see, for example, Ref. \cite{ovalle2018}) or with the BTZ one in $2+1$ dimensions
	(see Ref. \cite{contreras2019a} for details) leads to the mimic constraint
	of the radial pressure, namely $p=p^{\theta}_{r}$.
	
	\item Hairy Black Hole. In this case, as long as the gravitational source remains generic,
	we must impose extra constraints. This can be done by imposing, for example, either an equation of state or any other geometric restriction on the space-time. However, if the energy-momentum tensor is not generic, 
	perhaps the imposition of extra conditions
	is unnecessary. Indeed, this is the case 
	when the energy-momentum tensor is associated with a scalar 
	field. Here we choose to impose an equation of state for the components
	of $\theta^{\mu}_{\nu}$. Examples of such equations are the barotropic and perfect polytropic fluids. The
	reader is referred to Ref. \cite{ovalle2018a} for the $3+1$ dimensional case and to 
	Ref. \cite{contreras2018} for the $2+1$ case.
	
	\item Inverse problem. In this case, the constraint that allows the application of the method is 
	simply $\tilde{p}_{\perp}-\tilde{p}_{r}=p^{\theta}_{\perp}-p^{\theta}_{r}$, 
	where $\tilde{p}_{\perp},\tilde{p}_{r}$ corresponds to the components of $T^{tot}_{\mu\nu}$. 
	In the inverse problem, a solution of Eq. (\ref{einsorig}) is assumed to be known and the task is to explore 
	the decoupler sectors. This problem has been worked out in $3+1$ and $2+1$ dimensions
	in Refs. \cite{contreras2018a,contreras2018c,contreras2019}.
\end{itemize}

However, it is worth mentioning that the main limitation of the MGD approach 
is that the geometric deformation is performed only in the radial component of the metric. 
In order to overcome this limitation, an extension to the MGD method
has been proposed very recently in $3+1$ dimensions by Ovalle in Ref. \cite{ovalleplb}. In this work, the author 
introduced a modification in two components of the metric in a spherically symmetric space--time. The main 
result of that work is that the sources can be successfully decoupled
as long as there is exchange of energy between them. 
Following {Ref. \cite{ovalleplb}, in this work we obtain the extended 
	version of the MGD-decoupling in $2+1$ dimensions with cosmological constant.
	
	This work is organized as follows. In the next section we briefly review the MGD-decoupling method
	and we obtain its extension in $2+1$ dimensions with cosmological constant. 
	In section \ref{BTZM} we apply the extended method to study the generation of the BTZ-Maxwell solution starting from the
	BTZ vacuum. Section \ref{sch} is devoted to
	study the consequences of the Schwarzschild condition in the context of the extended gravitational decoupling. In section \ref{BHbaro} we use the decoupled system and implement barotropic and polytropic equations of state in the anisotropic sector to extend the BTZ vacuum to new $2+1$ black hole solutions. In sections \ref{RegPoly} and \ref{GenPoly}
	we demonstrate that the decoupled system can be used to obtain regular black hole solutions. 
	The last section is devoted to final comments and conclusions.
	
	\section{Extended Einstein Equations in $2+1$ space--time dimensions}\label{mgd}
	Recently, a geometric deformation in the two components of the metric in $3+1$ dimensions has been successfully performed
	in Ref. \cite{ovalleplb}. More precisely
	\begin{eqnarray}\label{decoupling}
	\nu&=&\xi + \alpha g\\
	e^{-\lambda}&=&e^{-\mu}+\alpha f.
	\end{eqnarray}
	In this case the method allows to decouple the sources $T^{(m)}_{\mu\nu}$ and $\theta_{\mu\nu}$
	of Eq. (\ref{einsorig}), usually with exchange of energy between them. 
	As a particular case, the decoupling can be obtained
	without exchange of energy when: i) $T^{(m)}_{\mu\nu}$ is a barotropic
	fluid whose equation of state is $\rho=-p$  and ii) for those regions where $T^{(m)}_{\mu\nu}=0$
	which could correspond, for example, to the case of a region $r>R$ filled by a source $\theta_{\mu\nu}$
	surrounding a self--gravitating system of radius $R$ and source $T^{(m)}_{\mu\nu}$.
	
	In this work, our main goal is to propose a decomposition in the whole geometric sector which could allow 
	us to decouple the system in $2+1$ dimensional space--times and explore the conditions under
	which the decoupling can be reached.
	
	Let us start by considering Eq. (\ref{le}) as a solution of the Einstein Field Equations, namely
	\begin{eqnarray}
	\kappa ^2 \tilde{\rho}&=&-\Lambda+\frac{e^{-\lambda} \lambda'}{2 r}\label{eins1}\\
	\kappa ^2 \tilde{p}_{r}&=&\Lambda+\frac{e^{-\lambda} \nu '}{2 r}\label{eins2}\\
	\kappa ^2 \tilde{p}_{\perp}&=&\Lambda+\frac{1}{4} e^{-\lambda} \left(-\lambda ' \nu '+2 \nu ''+\nu '^2\right)\label{eins3},
	\end{eqnarray}
	where the prime denotes derivation with respect to the radial coordinate and we have defined
	\begin{eqnarray}
	\tilde{\rho}&=&\rho+\rho^{\theta}\label{rot}\\
	\tilde{p}_{r}&=&p+p_{r}^{\theta}\label{prt}\\
	\tilde{p}_{\perp}&=&p+p_{\perp}^{\theta}.\label{ppt}
	\end{eqnarray} 
	
	The next step consists in decoupling the Einstein field
	equations (\ref{eins1}), (\ref{eins2}) and (\ref{eins3}) by performing the decomposition
	\begin{eqnarray}\label{def}
	&&\nu=\xi+\alpha g\\
	&&e^{-\lambda}=e^{-\mu} + \alpha f.
	\end{eqnarray}
	By doing so, we obtain
	two sets of differential equations: one describing a system sourced by
	the conserved energy--momentum tensor of a fluid $T_{\mu\nu}^{(m)}$ (which is conserved with respect to the 
	isotropic metric) and the other
	set corresponding to the equation of motion of the source $\theta_{\mu\nu}$. 
	After taking into account that the cosmological constant can be interpreted as an isotropic fluid, 
	we include the $\Lambda$--term in the isotropic sector and
	we obtain
	
	\begin{eqnarray}
	\kappa^{2}\rho &=&-\Lambda +\frac{e^{-\mu} \mu'}{2 r}\label{iso1}\\
	\kappa ^2 p&=&\Lambda +\frac{e^{-\mu} \xi'}{2 r}\label{iso2}\\
	\kappa ^2 p&=&\Lambda -\frac{e^{-\mu} \left(\mu' \xi'-2 \xi''-\xi'^2\right)}{4}
	,\label{iso3}
	\end{eqnarray}
	for $T_{\mu\nu}^{(m)}$ and
	\begin{eqnarray}
	\kappa ^2\rho^{\theta}&=&-\frac{\alpha  f'}{2 r}\label{aniso1}\\
	\kappa^{2} p_{r}^{\theta}&=&\alpha Z_{1}+\frac{\alpha f\nu'}{2 r}\label{aniso2}\\
	\kappa^{2} p_{\perp}^{\theta}&=&\alpha Z_{2}+\frac{\alpha}{4}f'\nu'+\frac{\alpha}{4 } 
	f\left(2 \nu ''+\nu '^2\right),\label{aniso3}
	\end{eqnarray}
	where
	\begin{eqnarray}
	Z_{1}&=&\frac{e^{-\mu} g'}{2 r}\\
	Z_{2}&=&\frac{1}{4} e^{-\mu} \left(2 g''+g' \left(\alpha  g'-\mu'+2 \xi'\right)\right)
	\end{eqnarray}
	for the source $\theta_{\mu\nu}$. 
	We would like to emphasize that the addition of the cosmological constant only affects 
	the isotropic sector because  Eqs. (\ref{aniso1}), (\ref{aniso2})
	and (\ref{aniso3}) remain unchanged. 
	
	In what follows we shall explore the Bianchi identities in order to study the
	conditions for the gravitational decoupling. 
	The conservation $\nabla_{\mu}T^{\mu(tot)}_{\nu}=0$ leads to
	\begin{equation}
	\left(p'+\frac{1}{2}(p+\rho) \xi'\right)+\frac{1}{2}\alpha g'(p+\rho)
	+\alpha  \left(\frac{1}{2} (p_{r}^{\theta}+\rho^{\theta})\nu '+\frac{-p_{\perp}^{\theta}+p_{r}^{\theta}+r p_{r}^{\theta}\ '}{r}\right)=0.
	\end{equation}
	In the above equation, the first bracketed term corresponds to the conservation of the radial component of $T^{(m)}_{\mu\nu}$ 
	computed with the metric $(\xi,\mu)$, namely 
	\begin{eqnarray}
	\nabla^{(\xi,\mu)}_{\rho}T^{\rho(m)}_{1}=p'+\frac{1}{2}(p+\rho) \xi'=0,
	\end{eqnarray}
	while, the third bracketed term
	corresponds to $\nabla_{\mu}\theta^{\mu}_{1}$ calculated with metric $(\nu,\lambda)$, namely
	\begin{eqnarray}
	\nabla_{\rho}\theta^{\rho}_{1}=p_{r}^{\theta}\ ' + \frac{1}{2} (p_{r}^{\theta}+\rho^{\theta})\nu '+\frac{p_{r}^{\theta}-p_{\perp}^{\theta}}{r}.
	\end{eqnarray}
	Now, with the previous notation, the conservation of the total energy momentum tensor can be written as
	\begin{eqnarray}
	\nabla_{\rho}T^{\rho(tot)}_{\nu}=\nabla^{(\xi,\mu)}_{\rho}T^{\rho(m)}_{\nu}+\nabla_{\rho}\theta^{\rho}_{\nu}
	+\frac{1}{2}\alpha g'(p+\rho)\delta_{\nu}^{1}=0.
	\end{eqnarray}
	Note that the total energy momentum tensor is conserved whenever 
	
	\begin{eqnarray}
	\nabla_{\rho}\theta^{\rho}_{\nu}=
	-\frac{1}{2}\alpha g'(p+\rho)\delta_{\nu}^{1},
	\end{eqnarray}
	which means decoupling with exchange of energy--momentum. However, a decoupling without energy--momentum exchange 
	can be reached either imposing $g'=0$ or $p+\rho=0$. The former requirement corresponds to the standard
	MGD where only $g^{-1}_{rr}$ undergoes a geometrical deformation. The latter
	entails a barotropic equation of state in the isotropic sector. What is more, if the isotropic 
	sector is vacuum (the exterior of a star), the barotropic condition is trivially fulfilled and
	the decoupling without exchange of energy--momentum is straightforward. We conclude this section
	pointing out that the conditions for the decoupling of the sources $T^{(m)}_{\mu\nu}$ and
	$\theta_{\mu\nu}$ coincides with those found for the $3+1$ case reported in Ref. {\cite{ovalleplb}}. 
	
	In the next section, we shall implement the extended MGD protocol in order to test it by generating an exterior
	charged BTZ solution starting form a BTZ vacuum.
	
	\section{BTZ-Maxwell system}\label{BTZM}
	Let us consider the static BTZ metric as the solution for the isotropic sector given by
	Eqs. (\ref{iso1}), (\ref{iso2}) and (\ref{iso3}),
	\begin{eqnarray}
	\xi &=& \log(-M+\frac{r^{2}}{L^{2}})\\
	\mu &=&-\log(-M+\frac{r^{2}}{L^{2}}).
	\end{eqnarray}
	The decoupler matter content and the Maxwell energy momentum tensor are given, respectively, by
	
	\begin{eqnarray}
	\theta^{\nu}_{\mu}=\frac{1}{4\pi}\left(F_{\mu\sigma}F^{\nu\sigma}-\frac{1}{4}\delta^{\nu}_{\mu}F_{\tau\sigma}F^{\tau\sigma}
	\right),
	\end{eqnarray}
	where $F_{\mu\nu}$ satisfies
	\begin{eqnarray}
	\nabla_{\nu}F^{\mu\nu}&=&4\pi j ^{\mu}\label{max}\\
	\partial_{[\sigma}F_{\mu\nu]}&=&0.
	\end{eqnarray}
	In the previous equations, $F_{\mu\nu}$ and  $j^{\mu}$ are the Maxwell tensor and 
	the four-current, respectively. Now, in the static and circularly symmetric case,
	the only non-vanishing components of the Maxwell tensor are $F^{01}=F^{10}$ and the four current
	reads
	\begin{eqnarray}
	j^{\mu}&=&(j^{0},0,0).
	\end{eqnarray}
	After integration of the Maxwell equation (see Eq. (\ref{max})) we obtain
	\begin{eqnarray}
	F^{01}=\frac{e^{-(\nu+\lambda)/2}q(r)}{r},
	\end{eqnarray}
	where $q(r)$ is the electric charge of a spherical system of radius $r$, defined as
	\begin{eqnarray}
	q(r)=\int\limits^{r}4\pi  e^{(\nu+\lambda)/2}j^{0}RdR.
	\end{eqnarray}
	With these results at hand, Eqs. (\ref{aniso1}), (\ref{aniso2}) and (\ref{aniso3}) become
	\begin{eqnarray}
	-E^{2}&=&-\frac{\alpha f'}{2 r}\label{ani1}\\
	E^{2}&=&\alpha Z_{1}+\frac{\alpha f\nu'}{2 r}\label{ani2}\\
	-E^{2}&=&\alpha Z_{2}+\frac{\alpha}{4}f'\nu+\frac{\alpha}{4} f\left(2 \nu ''+\nu '^2\right)\label{ani3}.
	\end{eqnarray}
	From the conservation equation, the electric field turns to be
	\begin{eqnarray}\label{E}
	E=\frac{Q}{r},
	\end{eqnarray}
	where $Q$ is the total charge of the black hole. After substituting Eq. (\ref{E})
	in (\ref{ani1}), (\ref{ani2}) and (\ref{ani3}) we obtain
	\begin{eqnarray}
	f(r)&=& c_1+\frac{2 Q^2 \log (r)}{\alpha }\\
	g(r)&=&\frac{1}{\alpha }\log \left(\frac{L^2 \left(\alpha  c_1-M\right)+2 L^2 Q^2 \log (r)+r^2}{r^2-L^2 M}\right)+c_2.
	\end{eqnarray}
	Now, using the decoupling constraints given by Eqs. (\ref{def}), we obtain
	\begin{eqnarray}
	\nu&=&\alpha  c_2+\log \left(L^2 \left(\alpha  c_1-M\right)+2 L^2 Q^2 \log (r)+r^2\right)\nonumber\\
	&&+\log \left(\frac{r^2}{L^2}-M\right)-\log \left(r^2-L^2 M\right)\\
	\lambda &=&-\log \left(\alpha  c_1+\frac{r^2}{L^2}-M+2 Q^2 \log (r)\right),
	\end{eqnarray}
	which, after some manipulation can be written as
	\begin{eqnarray}
	e^\nu &=& -M+\frac{r^2}{L^2}+2 Q^2 \log \left(\frac{r}{L}\right)\\
	e^{-\lambda}&=&-M+\frac{r^2}{L^2}+2 Q^2 \log \left(\frac{r}{L}\right),
	\end{eqnarray}
	which corresponds to the well know charged BTZ black hole solution. 
	
	To conclude this section, a couple of comments are in order. First, it is worth mentioning that we have 
	extended our previous work concerning the MGD approach in 2+1 dimensions \cite{contreras2018} to the case of two deformations instead of one. 
	In this sense, more complex situations can be tackled in three--dimensional space--times with the improvement here presented.
	Second, we note that we have been able to reconstruct the charged--BTZ solution starting from the corresponding vacuum, which is
	pure BTZ. In this sense, our findings are equivalent to that of the $3+1$--dimensional case presented in 
	Ref. \cite{ovalleplb}. 
	In the next section we propose a particular parametrization for a circularly symmetric metric which, not only allows to simplify the problem but leads to the decoupling of the sources involved.
	
	\section{Extended-MGD and Schwarzschild condition}\label{sch}

	In this section we establish a relationship between the radial and the temporal deformation functions in the 
	extended-MGD program with the aim that the metric of the total solution satisfies the Schwarzschild condition, namely 
	$g_{tt}=g_{rr}^{-1}$. As it will be seen later, this special requirement leads to a noticeable simplification of the decoupled 
	equations.
	
	As a first step, we
	parametrize the metric of the  isotropic sector by
	\begin{eqnarray}\label{isoF}
	e^{\xi}=e^{-\mu}=F(r)
	\end{eqnarray}
	and define the radial decoupling function $g(r)$ as
	\begin{eqnarray}\label{gr}
	g(r)=\frac{1}{\alpha }\log \left(\frac{\alpha  f+F}{F}\right).
	\end{eqnarray}
	Finally, by replacing Eqs. (\ref{isoF}) and (\ref{gr}) in (\ref{decoupling}), the
	the metric potentials of the line element of the total solution, $ds^{2}=-e^{\nu}dt^{2}+e^{\lambda}dr^{2}+r^{2}d\phi^{2}$, read
	\begin{eqnarray}
	e^{\nu}=e^{-\lambda}=F+\alpha f,
	\end{eqnarray}
	which satisfy the Schwarzschild condition, as required. Furthermore, the
	system of differential equations (\ref{iso1}), (\ref{iso2}), (\ref{iso3}), (\ref{aniso1}), (\ref{aniso2}) and (\ref{aniso3}) reduces 
	to
	
	\begin{eqnarray}
	\rho&=&\frac{\Lambda  }{8\pi } -\frac{F'}{16\pi r}\label{ddec} \\
	p_{r} &=&\frac{\Lambda  }{8\pi }+ \frac{F'}{16\pi r}\label{prdec}\\
	p_{\perp}&=&\frac{ \Lambda }{ 8\pi}+\frac{F''}{16\pi}\label{ptdec},
	\end{eqnarray}
	for the isotropic sector, with the condition $p_{r}=p_{\perp}$, while for the source $\theta_{\mu\nu}$ we have
	\begin{eqnarray}
	\rho^{\theta}&=&-\frac{\alpha  f'}{16 \pi  r}\label{thedec}\\
	p_{r}^{\theta} &=& \frac{\alpha  f'}{16 \pi  r}\label{pdec}\\
	p_{\perp}^{\theta} &=& \frac{\alpha  f''}{16 \pi }\label{ppdec}.
	\end{eqnarray}
	Note that, as commented before, this new set of differential equations corresponds to a considerably simplified 
	version of the original one which, as we will show in the next sections, allows to obtain new solutions for the decoupled system 
	if we use, for example, the static BTZ black hole with $F=-M+\frac{r^{2}}{L^{2}}$ as the isotopic sector and some appropriate 
	equation of state for the components of $\theta_{\mu\nu}$.
	\section{Black hole solution by a barotropic equation of state}\label{BHbaro}
	In order to find a solution for $f$ we proceed to implement an appropriate equation of state. For example, 
	for the barotropic equation of state
	\begin{eqnarray}
	\rho^{\theta}=A p^{\theta}_{r} +B p_{\perp}^{\theta},
	\end{eqnarray}
	the solution reads
	\begin{eqnarray}
	f=\frac{c_1 }{1-\frac{A+1}{B}}r^{1-\frac{A+1}{B}}.
	\end{eqnarray}
	The above solution can be used to construct a new anisotropic solution with isotropic sector parametrized 
	as $ds^{2}=-F dt^{2}+\frac{1}{F}dt^{2}+r^{2}d\phi^{2}$. For example, for the static BTZ solution we have 
	$F=-M+\frac{r^{2}}{L^{2}}$, the solution to Eqs. (\ref{eins1}), (\ref{eins2}) and (\ref{eins3}) are given by
	\begin{eqnarray}
	e^{\nu}=e^{-\lambda}=-M+\frac{r^2}{L^2}- \alpha\frac{B c_1 }{A-B+1}r^{\frac{-A+B-1}{B}}.
	\end{eqnarray}
	
	The above corresponds to a new family of anisotropic solution in $2+1$ dimensions with sources 
	given by
	\begin{eqnarray}
	\tilde{\rho}&=& -\frac{\alpha  c_1 r^{-\frac{A+B+1}{B}}}{16 \pi }\\
	\tilde{p}_{r}&=& \frac{\alpha  c_1 r^{-\frac{A+B+1}{B}}}{16 \pi }\\
	\tilde{p}_{\perp}&=& -\frac{\alpha  (A+1) c_1 r^{-\frac{A+B+1}{B}}}{16 \pi  B}.
	\end{eqnarray}
	The invariant scalars are given by 
	\begin{eqnarray}
	R&=&\frac{\alpha  c_1 (A-2 B+1) r^{-\frac{A+B+1}{B}}}{B}-\frac{6}{L^2}\\
	Ricc^{2}&=&\frac{\alpha ^2 c_1^2 \left(-2 (A+1) B+(A+1)^2+3 B^2\right) r^{-\frac{2 (A+B+1)}{B}}}{2 B^2}\nonumber\\
	&&-\frac{4 \alpha  c_1 (A-2 B+1) r^{-\frac{A+B+1}{B}}}{B L^2}+\frac{12}{L^4}\\
	\mathcal{K}&=&\frac{\alpha ^2 c_1^2 \left((A+1)^2+2 B^2\right) r^{-\frac{2 (A+B+1)}{B}}}{B^2}\nonumber\\
	&&-\frac{4 \alpha  c_1 (A-2 B+1) r^{-\frac{A+B+1}{B}}}{B L^2}+\frac{12}{L^4},
	\end{eqnarray}
	which correspond to the Ricci, the Ricci squared and the Kretschmann scalar respectively and clearly differ form those of 
	the static BTZ solution.
	
	The horizons of the solution, $r_{H}$, correspond to
	the real roots of $e^{-\lambda(r_{H})}=0$. More precisely
	\begin{eqnarray}
	-\frac{\alpha B c_1 }{A-B+1}r_H^{\frac{-A+B-1}{B}}+\frac{r_H^2}{L^2}-M=0.
	\end{eqnarray}
	
	It is worth noticing that after an appropriate choice of the constants, the solution looks like
	the $4$-dimensional Schwarschild-AdS solution but this time in a $3$--dimensional spacetime. Indeed, taking $c_{2}=(1+M)/\alpha$, $A=B=1$ and
	$c_{1}=2M/\alpha$ we obtain
	\begin{eqnarray}
	e^{\nu}=e^{-\lambda}=1-\frac{2M}{r}+\frac{r^{2}}{L^{2}}
	\end{eqnarray}
	
	In this case, the invariants read
	\begin{eqnarray}
	R&=&-\frac{6}{L^{2}}\\
	Ricc^{2}&=&\frac{12}{L^4}+\frac{6 M^2}{r^6}\\
	\mathcal{K}&=&\frac{12}{L^4}+\frac{24 M^2}{r^6}
	\end{eqnarray}
	and the solution is sourced by
	\begin{eqnarray}
	\tilde{\rho}&=&-\frac{M}{8 \pi  r^3}\\
	\tilde{p}_{r}&=&\frac{M}{8 \pi  r^3}\\
	\tilde{p}_{\perp}&=&-\frac{M}{4 \pi  r^3}
	\end{eqnarray}
	Note that, this matter sector violates all the energy conditions. Although, in general, this result
	could be taken as problematic, it could be of importance in the
	context of the inverse problem, which allowed to interpret the MGD as some kind of mechanism which leads to the apparition of 
	exotic matter after gravitational interaction of well behaved matter content \cite{contreras2018a}.
	
	\section{Polytropic equation of state of the form $p_{\perp}=k_{0}(\rho^{\theta})^{2}$. A regular black hole solution }\label{RegPoly}
	In this section we obtain  regular black hole solution by considering a equation of state of the form
	\begin{eqnarray}
	p^{\theta}_{\perp}=k_{0}(\rho^{\theta})^{2}.
	\end{eqnarray}
	Replacing the above condition in Eqs. (\ref{thedec}) , (\ref{pdec}) and (\ref{ppdec}) and solving the differential equation for the decoupling function we obtain
	\begin{eqnarray}
	f=-\frac{r}{c_1}-\frac{\alpha k_{0} \log \left(\alpha  k_{0}-16 \pi  c_1 r\right)}{16 \pi  c_1^2}.
	\end{eqnarray}
	In this case, combining the above solution with Eqs. (\ref{isoF}) and (\ref{gr}), the metric functions of the total solution (Eqs. (\ref{def})) are given by
	\begin{eqnarray}\label{regu}
	F&=&-M+\frac{r^2}{L^2}
	-\frac{\alpha r}{c_1}   -\alpha\frac{\alpha  k_{0} \log \left(\alpha  k_{0}-16 \pi  c_1 r\right)}{16 \pi  c_1^2}.
	\end{eqnarray}
	The scalars for this solutions are given by
	\begin{eqnarray}
	R&=&\frac{16 \pi  \alpha  \left(32 \pi  c_1 r-3 \alpha  k_{0}\right)}{\left(\alpha k_{0}
		-16 \pi  c_1 r\right){}^2}-\frac{6}{L^2}\\
	Ricc^{2}&=&\frac{12}{L^4}+\frac{256 \pi ^2 \alpha ^2 \left(-64 \pi  \alpha  c_1 k_{0} r+384 \pi ^2 c_1^2 r^2+3 \alpha ^2 k_{0}^2\right)}{\left(\alpha k_{0}-16 \pi  c_1 r\right){}^4}\nonumber\\
	&&+\frac{64 \pi  \alpha  \left(3 \alpha  k_{0}-32 \pi  c_1 r\right)}{L^2 \left(\alpha  k_{0}-16 \pi  c_1 r\right){}^2}\\
	\mathcal{K}&=&\frac{12}{L^4}+\frac{64 \pi  \alpha  \left(8 \pi  \left(\alpha  L^2-4 c_1 r\right)+3 \alpha  k_{0}\right)}{L^2 \left(\alpha  k_{0}-16 \pi  c_1 r\right){}^2}
	+\frac{256 \pi ^2 \alpha ^4 k_{0}^2}{\left(\alpha  k_{0}-16 \pi  c_1 r\right){}^4},
	\end{eqnarray}
	and the sources can be written as
	\begin{eqnarray}
	\tilde{\rho}&=&\frac{\alpha }{16 \pi  c_1 r-\alpha  k_{0}}\label{rho0}\\
	\tilde{p}_{r}&=&\frac{\alpha }{\alpha  k_{0}-16 \pi  c_1 r}\label{pr0}\\
	\tilde{p}_{\perp}&=&\frac{\alpha ^2 k_{0}}{\left(\alpha  k_{0}-16 \pi  c_1 r\right){}^2}\label{pp0}.
	\end{eqnarray}
	At this point some comments are in order. First, the solution corresponds to a regular black hole whenever $c_{1}<0$. However, this condition would implies a negative energy density unless $k_{0}<0$ and $\alpha<0$. What is more, with the above conditions the solution fulfils the weak energy condition, namely
	\begin{eqnarray}
	&&\tilde{\rho}\ge0\\
	&&\tilde{\rho}+\tilde{p}_{r}\ge 0\\
	&&\tilde{\rho}+\tilde{p}_{\perp}\ge 0.
	\end{eqnarray}
	Second, the horizons of this solution corresponds to the real roots of
	\begin{eqnarray}
	e^{-\lambda(r_{H})}&=&-\frac{\alpha^{2}  k_{0} \log \left(\alpha  k_{0}-16 \pi  c_1 r_{H}\right)}{16 \pi  c_1^2}
	-\frac{\alpha r_{H}}{c_1}+\frac{r_{H}^2}{L^2}-M=0.
	\end{eqnarray}
	In figure \ref{fig1} we show the behaviour of $F(r)$ for 
	$M=1$, $L=3$, $ k_{0} =-10$, $ c_{1}= -0.5$ and different values of $\alpha$. Note the appearance of
	two horizons as in the standard 2+1 and 3+1 charged solutions. In this sense, it could be interesting to
	study if there is also a perturbative instability of the inner horizon as in the charged BTZ solution \cite{Martinez2000}.
	\begin{figure}[h!]
		\centering
		\includegraphics[scale=0.6]{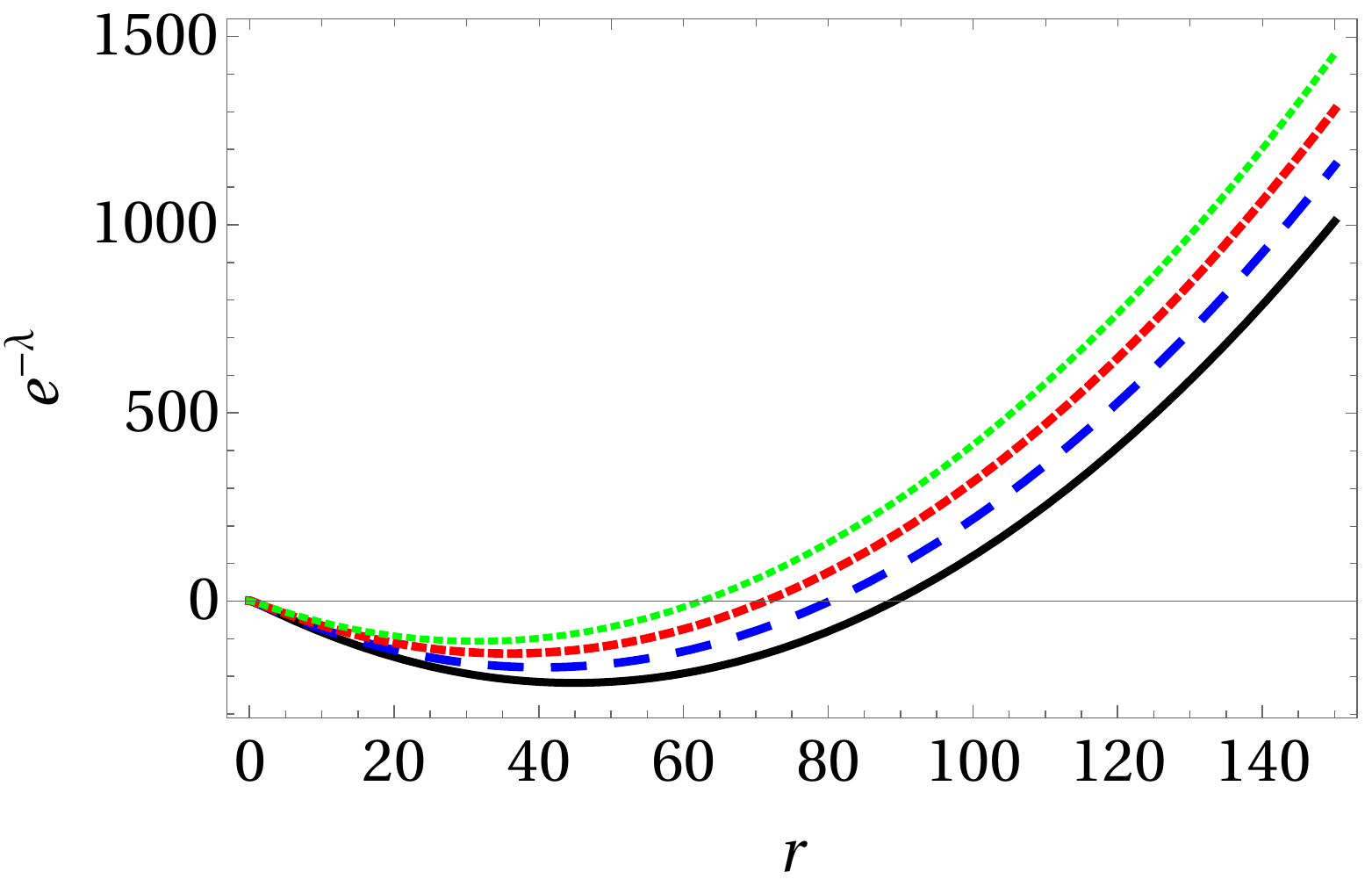}
		\caption{\label{fig1} 
			Metric function for for $\alpha =-10$ (black solid line), $\alpha=-9$ (dashed blue line)
			$\alpha =-8$ (short dashed red line) and $\alpha =-7$ (dotted green line).
		}
	\end{figure}
	It is worth mentioning that the regular black hole solution Eq. (\ref{regu}) looks formally like
	the regular charged solution (see for example \cite{He2017})
	\begin{equation}\label{NL}
	ds^{2}=-g dt^{2}+g^{-1}dr^{2}+r^{2}dt^{2},
	\end{equation}
	where
	\begin{eqnarray}
	g=-M+\frac{r^{2}}{L^{2}}-q^{2}\log \left( \frac{q^{2}/a^{2}+r^{2}}{L^{2}}  \right).
	\end{eqnarray}
	However, the main difference between them is the term proportional to $r$ appearing in our solution (see Eq. (\ref{regu})).
	
	Before concluding this section, let us briefly discuss a possible interpretation in terms of non--linear electrodynamics associated to
	our solution obtained by the extended MGD. As it is well known (see for example \cite{garcialibro}), in $2+1$
	dimensions, the Einstein-Non--linear electrodynamics system is sourced by
	\begin{eqnarray}
	\tilde{\rho}&=&g(L+E^{2}L_{,F})\label{rhoNL}\\
	\tilde{p}_{r}&=&\frac{L+E^{2}L_{,F}}{g}\label{prNL}\\
	\tilde{p}_{\perp}&=&r^{2}L\label{ppNL},
	\end{eqnarray}
	where $g(r)$ is the metric function of (\ref{NL}), $E(r)$ is the electric field, $L(r)$ is the non--linear electrodynamics Lagrangian and
	$L_{,F}=dL/dF$ where $F=F_{\mu\nu}F^{\mu\nu}$ with $F_{\mu\nu}$ the usual Maxwell tensor. An extra constraint arises from the non--linear Maxwell equation, namely
	\begin{eqnarray}
	L_{,F}=-\frac{q}{r E },
	\end{eqnarray}
	where $q$ is a constant with dimensions of electric charge. In our case, if we identify
	the metric function in Eq. (\ref{NL}) with $g=e^{\nu}=e^{-\lambda}=F+\alpha f$, and the set of Eqs. (\ref{rhoNL}), (\ref{prNL})
	and (\ref{ppNL}) with Eqs. (\ref{rho0}), (\ref{pr0}) and (\ref{pp0}) we obtain that the non--linear electrodynamics associated to
	the extended MGD with polytropic equation of state is characterized by
	\begin{eqnarray}
	E&=&\frac{\alpha ^2 k_{0}}{q r \left(\alpha  k_{0}-16 \pi  c_1 r\right){}^2}\\
	L&=&\frac{\alpha ^2 k_{0}}{r^2 \left(\alpha  k_{0}-16 \pi  c_1 r\right){}^2}.
	\end{eqnarray}
	
	At this point some comments are in order. First, the above solution corresponds to the electric field and to
	the Lagrangian 
	associated to the anisotropic sector of the decoupled extended MGD. Second, in contrast to the 
	standard procedure to construct regular solutions which consists in proposing a suitable metric function or a regular electric field, in this case the protocol consists in imposing a suitable equation of state. It is worth noticing that the former procedure is an alternative to what is usually found in the literature, which consists in the imposition of certain distributions to regularize the geometry that seems to be slightly artificial. Third, note that although
	the geometry is regular at the origin, the non--linear Lagrangian and its associated electric field
	diverge when $r$ goes to zero. A similar situation can be found in Ref. \cite{cataldo}, but in this
	case, given that electrodynamic source corresponds to a Born--Infeld model, the sources 
	are regular everywhere but the scalars diverge at $r=0$. Finally, we think that the extended MGD, 
	interpreted in its most general case without entering in any particular solution in terms of non-linear electrodynamics, 
	could serve to construct regular black hole solutions and maybe to associate their regularity with the geometric deformations 
	which are one of the main features of the MGD method.

	\section{General polytropic equation of state}\label{GenPoly}
	For a general polytropic equation of state of the form $p^{\theta}_{\perp}=k_{0}\rho^{A}$, the decoupling function $f$ can be written as
	\begin{eqnarray}
	f&=&c_2+r \left(1+\frac{e^{i \pi  A} (16 \pi )^{1-A} k_{0} \alpha ^{A-1} r^{1-A}}{c_1-A c_1}\right)\times\nonumber\\
	&&\times\left(-A c_1+e^{i \pi  A} (16 \pi )^{1-A} k_{0} \alpha ^{A-1} r^{1-A}+c_1\right){}^{\frac{1}{1-A}} \,\times\nonumber\\
	&&\times _2F_1\left(1,\frac{A-3}{A-1};1+\frac{1}{1-A};\frac{e^{i A \pi } k_{0} (16 \pi )^{1-A} r^{1-A} \alpha ^{A-1}}{(A-1) c_1}\right),
	\end{eqnarray}
	where $_{2}F_{1}(a;b;c;z)$ corresponds to the hypergeometric function. It is worth mentioning that the case studied in the previous section can not be obtained by choosing $A=2$. Indeed, $A=2$ leads to a complex infinity value of $f$ in this case. 
	
	As a special case, we could consider $\gamma=3$ from where
	\begin{eqnarray}
	f=-\frac{r \sqrt{-512 \pi ^2 c_1-\frac{\alpha ^2 k_{0}}{r^2}}}{32 \pi  c_1}.
	\end{eqnarray}
	Note that, in order to obtain a real solution, we must impose $c_{1}<0$ and $k_{0}<0$. In terms of this
	decoupling metric, the total solution reads
	\begin{eqnarray}
	e^{\nu}=e^{-\lambda}=-M+\frac{r^2}{L^2}-\alpha \frac{r \sqrt{512 \pi ^2 |c_1|
			+\frac{\alpha ^2 |k_{0}|}{r^2}}}{32 \pi  c_1}.
	\end{eqnarray}
	The invariants associated to this solution reveal that it is regular everywhere. Indeed,
	\begin{eqnarray}
	R&=& -\frac{6}{L^2}+\frac{16 \pi  \alpha  r  \left(1024 \pi ^2 c_1 r^2+3 \alpha ^2 k_{0}\right)}{\left(512 \pi ^2 c_1 r^2+\alpha ^2 k_{0}\right){}^2}\mathcal{F}(r)\\
	Ricc^{2}&=& -\frac{64 \pi  \alpha  r  \left(1024 \pi ^2 c_1 r^2+3 \alpha ^2 k_{0}\right)\mathcal{F}(r)}{L^2 \left(512 \pi ^2 c_1 r^2+\alpha ^2 k_{0}\right){}^2}\nonumber\\
	&&-\frac{256 (2048 \pi ^4 \alpha ^4 c_1 k_{0} r^2+393216 \pi ^6 \alpha ^2 c_1^2 r^4)}{\left(512 \pi ^2 c_1 r^2+\alpha ^2 k_{0}\right){}^3}\nonumber\\
	&&+\frac{256\times 3 \pi ^2 \alpha ^6 k_{0}^2}{\left(512 \pi ^2 c_1 r^2+\alpha ^2 k_{0}\right){}^3}+\frac{12}{L^4}\\
	\mathcal{K}&=& \frac{\mathcal{G}(r)}{\left(512 \pi ^2 c_1 r^2+\alpha ^2 k_{0}\right){}^3},
	\end{eqnarray}
	where
	\begin{eqnarray}
	\mathcal{F}=\sqrt{512 \pi ^2 |c_1|+\frac{\alpha ^2 |k_{0}|}{r^2}},
	\end{eqnarray}
	and $G(r)$ is a complicated expression involving powers of $r$ and the function $\mathcal{F}$. The horizons of the solution are located at
	\begin{eqnarray}
	r_{\pm}=\sqrt{\frac{8 \pi  |c_1| L^2 \left(\alpha ^2 L^2+4 |c_1| M\right)\pm H(r)}{32 \pi  c_1^2}},
	\end{eqnarray}
	where
	\begin{eqnarray}
	H=\sqrt{\alpha ^2 c_1^2 L^4 \left(64 \pi ^2 L^2 \left(\alpha ^2 L^2+8 |c_1| M\right)+\alpha ^2 
		|k_{0}|\right)}.
	\end{eqnarray}
	The sources of this solution are given by
	\begin{eqnarray}
	\tilde{\rho}&=&-\frac{\alpha }{\sqrt{512 \pi ^2 |c_1| r^2+\alpha ^2 |k_{0}|}}\\
	\tilde{p}_{r}&=&\frac{\alpha }{\sqrt{512 \pi ^2 |c_1| r^2+\alpha ^2 |k_{0}|}}\\
	\tilde{p}_{\perp}&=&\frac{\alpha ^3 |k_{0}|}{\left(512 \pi ^2 |c_1| r^2+\alpha ^2 |k_{0}|\right){}^{3/2}}.
	\end{eqnarray}
	Note that, in order to have a positive energy density an extra condition arises, namely $\alpha<0$. Even more, the 
	solution satisfies the weak energy condition.
	
	\section{Conclusions}\label{remarks}
	In this work we have successfully extended the Minimal Geometric Deformation method in circularly symmetric 
	$2+1$--dimensional space--times with cosmological constant. This extended method allows the introduction of deformations 
	in two of the components of the metric tensor, which results in the decoupling of the sources of the Einstein field equations. 
	We found that this decoupling could be obtained without exchange of energy between the sources as far as the perfect fluid satisfies 
	a barotropic equation of state or in situations where the isotropic sector corresponds to a vacuum solution. As an example, 
	we have implemented the extended protocol to generate an exterior solution starting from a static BTZ vacuum resulting in a 
	charged--BTZ system, in agreement with its corresponding $3+1$--dimensional counterpart recently obtained in 
	Ref. \cite{ovalleplb} in the sense that the vacuum sector leads to a charged one if the
	anisotropic source is the  Maxwell energy--momentum tensor.
	
	We remarked that the extended method can 
	be applied to obtain new solutions after an appropriate choice for the source is done but, in this case, the implementation of 
	a suitable equation of state is an option. In any case, we have to provide an additional equation to close the system. However, even in the simplest case, obtaining solutions to the equations 
	is a difficult task. In order to overcome this problem, we demonstrated that after a suitable parametrization of the line element of each sector, the system can be decoupled and the steps to obtain new solutions are considerably simplified. As a particular application we obtained new $2+1$ solutions extending the well known BTZ vacuum and we illustrated how the method can be used to contruct regular black hole solutions.

	\section*{Acknowledgments}
	The author P. B. was supported by the Faculty of Science and Vicerrector\'{\i}a de Investigaciones of Universidad de los Andes, Bogot\'a, Colombia.
	\\
	\\

\end{document}